\documentclass[12pt,times]{article}
\title{Active learning of constitutive relation from mesoscopic dynamics for macroscopic modeling of non-Newtonian flows\footnote{The authors wish it to be known that the first two authors contributed equally to this work and thus, in their opinion, should be regarded as co-first authors.}}

\author{Lifei Zhao$^{1,2}$, Zhen Li$^{2}$\footnote{Corresponding Authors. Email: \href{mailto:zhen_li@brown.edu}{zhen\_li@brown.edu}~(Zhen Li)}, Bruce Caswell$^3$, Jie Ouyang$^{1}$\footnote{Corresponding Authors. Email: \href{mailto:jieouyang@nwpu.edu.cn}{jieouyang@nwpu.edu.cn}~(Jie Ouyang)} and George Em Karniadakis$^2$\footnote{Email: \href{mailto:george_karniadakis@brown.edu}{george\_karniadakis@brown.edu}~(George Em Karniadakis)} \\
\small{$^1$~Department of Applied Mathematics, Northwestern Polytechnical University, Xi'An 710129, China}\\
\small{$^2$~Division of Applied Mathematics, Brown University, Providence, RI 02912, USA}\\
\small{$^3$~School of Engineering, Brown University, Providence, RI 02912, USA} }

\date{\today}

\RequirePackage{lineno}
\usepackage[colorlinks=true]{hyperref}
\usepackage{graphicx}
\usepackage{multirow,array,color,mathrsfs,bm,subcaption}
\usepackage{epstopdf} % make WinEdt able to compile eps file by using the default compile button!

\usepackage{caption}
\DeclareGraphicsExtensions{.eps,.mps,.pdf,.jpg,.png}
\graphicspath{{figures/}{../figures/}}

\usepackage{fancyhdr}

\usepackage{amsmath}
\usepackage{amsfonts}

\usepackage{dsfont}

\usepackage[top=1.5cm,bottom=2.5cm,left=1.5cm,right=1.5cm]{geometry}

\usepackage[figuresright]{rotating}
\usepackage{extarrows}

\usepackage{fixltx2e}
\usepackage{xspace}

\usepackage{amsmath,amssymb}
\begin{document}

\maketitle

\begin{abstract}
We simulate complex fluids by means of an on-the-fly coupling of the bulk rheology to the underlying microstructure dynamics. In particular, a macroscopic continuum model of polymeric fluids is constructed without a pre-specified constitutive relation, but instead it is actively learned from mesoscopic simulations where the dynamics of polymer chains is explicitly computed. To couple the macroscopic rheology of polymeric fluids and the microscale dynamics of polymer chains, the continuum approach (based on the finite volume method) provides the transient flow field as inputs for the (mesoscopic) dissipative particle dynamics (DPD), and in turn DPD returns an effective constitutive relation to close the continuum equations. In this multiscale modeling procedure, we employ an active learning strategy based on Gaussian process regression (GPR) to minimize the number of expensive DPD simulations, where adaptively selected DPD simulations are performed only as necessary. Numerical experiments are carried out for flow past a circular cylinder of a non-Newtonian fluid, modeled at the mesoscopic level by bead-spring chains. The results show that only five DPD simulations are required to achieve an effective closure of the continuum equations at Reynolds number $Re=10$. Furthermore, when $Re$ is increased to 100, only one additional DPD simulation is required for constructing an extended GPR-informed model closure. Compared to traditional message-passing multiscale approaches, applying an active learning scheme to multiscale modeling of non-Newtonian fluids can significantly increase the computational efficiency. Although the method demonstrated here obtains only a local viscosity from the mesoscopic model, it can be extended to other multiscale models of complex fluids whose macro-rheology is unknown.
\end{abstract}

\section{Introduction}
The main motivation for investigating non-Newtonian fluids is that no fluid is virtually Newtonian except some simple fluids such as air and water~\cite{2001Balmforth}. The distinguishing feature of non-Newtonian fluids is that the microstructures present in variable rheological conditions are not only transient but can be easily changed by the application of low stresses~\cite{2010Chhabra}. Consequently, non-Newtonian fluids usually have complex stress-strain rate relationships and their apparent viscosities depend on the transient shear rate. In particular, polymeric fluids are  the most widely used industrial non-Newtonian fluids. Their non-Newtonian behavior stems from the chain-like molecular structure of polymers, whose length can be so long that the collective effect of structural reorganization of polymer chains affects the macroscale rheological properties~\cite{2014Huber}. Moreover, because the forces generated by the polymer relaxation depend on its original orientation, polymeric fluids can also exhibit significant memory effects,
i.e., the stress tensor depends on the strain history~\cite{1983Crochet}. As a result, unlike the linear constitutive relation of Newtonian fluids, the constitutive equation of non-Newtonian fluids becomes much more complicated~\cite{1995Bird}, which makes the modeling of non-Newtonian fluids a challenging problem.

In continuum approaches, an expression for the stress tensor in terms of various kinematic tensors is needed in the momentum equation and also in the energy equation~\cite{2015Larson}. This problem is similar to that arising in modeling turbulent flows~\cite{2016Wallace}, where an empirical model is needed for computing the Reynolds stress tensor for subgrid contributions to shear stress. Under the continuum hypothesis, various phenomenological models for the constitutive relation of non-Newtonian fluids have been developed to close the continuum equations. Examples include the power-law and Bingham plastics models~\cite{2011Chai}, the Oldroyd-B model~\cite{1950Oldroyd}, the finitely extensible non-linear elastic-Peterlin (FENE-P) model~\cite{1997Herrchen,2003Ma} and the Phan-Thien-Tanner model~\cite{1977Phan}, to name but a few.

Since the non-Newtonian macroscopic properties are strongly related to the dynamics of underlying microscopic structures, it is a straightforward idea to couple the macroscopic rheology of non-Newtonian fluids and the microscopic physics using multiscale simulations, in which a continuum approach is used to model the macroscopic behavior of polymeric fluids, while a microscopic model is used to describe the dynamics of underlying microstructures. Many research efforts have been devoted to coupling continuum equations to micro/meso-scale simulations. Bell et al.~\cite{1997Bell} combined a spectral method and Brownian dynamics to investigate the recovery of polymeric fluids after the cessation of shear flow. Wagner and Liu~\cite{2003Wagner} coupled continuum finite elements to molecular dynamics (MD) simulations, and tested their scheme on a one-dimensional lattice; Kojic et al.~\cite{2013Kojic} then extended it by coupling macroscale finite elements to dissipative particle dynamics (DPD) simulations for simple fluids. Using the idea of domain decomposition, Fedosov and Karniadakis~\cite{2009Fedosov} developed a hybrid multiscale method (triple-decker) to concurrently couple atomistic-mesoscopic-continuum models, and Li et al.~\cite{2015YLi} coupled DPD simulations and a macroscale finite element method for the Couette flows of polymer solutions. Also, Moreno et al.~\cite{2013Moreno} coupled a finite element model with smoothed DPD to capture the non-Newtonian behavior of blood flowing through arteries. Recently, Barnes et al.~\cite{2017Barnes} constructed an effective equation of state for a macroscopic finite element model from mesoscopic DPD simulations.

In general, there exist three main categories of multiscale approaches, namely sequential approaches, concurrent couplings and adaptive resolution schemes~\cite{2009Horstemeyer,2008Praprotnik}. In concurrent coupling and adaptive resolution schemes the time step is limited by the time step used for microscopic simulations, while the sequential approach also known as message-passing is more suitable for multiscale problems with apparent scale separation between macroscopic and microscopic systems~\cite{2009Horstemeyer}. In the present work, we consider the continuum system to be much larger than its microscopic counterpart, so that a macroscale element is large enough to contain a representative sample of the micro-system. Specifically, the macroscale continuum equations are solved by the finite volume method (FVM) whose benefits are both high computational efficiency and numerical stability. The polymer fluid is modeled by bead-spring chains whose coarse-grained dynamics are computed by DPD -- a method well suited for modeling mesoscopic phenomena with much greater efficiency than all-atom molecular dynamics~\cite{2013Saunders,2013Mills,2017Espanol,2017ZLi-DPD}.

In the multiscale coupling procedure, the continuum model of polymeric fluids will be constructed without a closed form of the constitutive model, which can be computed by performing DPD simulations where the dynamics of polymer chains is explicitly simulated. In this  proof-of-concept study, we assume that the fluid is inelastic and can be descried as a generalized Newtonian fluid at macroscale level, so that the stress field cam be computed from a steady shear flow in the DPD system. We note, however, that there are more rheological properties of non-Newtonian fluids beyond the shear stress that can be obtained from DPD simulations, such as the normal stress differences and the spectrum of relaxation times that can be used for modeling elasticity and fading memory in continuum approaches~\cite{1987Bird}. The on-the-fly communications between macroscale and mesoscopic solvers can be implemented {\em seamlessly} by using a multiscale universal interface (MUI) library~\cite{2015Tang}. In general, to obtain an accurate function of the non-Newtonian viscosity in terms of shear rate requires many DPD simulations~\cite{2010Fedosov}, making this process computationally prohibitive as  DPD simulations of polymer models are expensive. Hence, in this work we will mitigate this computational expense by employing an {\em active learning} strategy together with the Gaussian process regression (GPR) to obtain the fluid's non-Newtonian viscosity only when necessary. Gaussian process models are particularly useful for regression because they provide not only the mean function response but also the corresponding uncertainty, which naturally allows for an active learning paradigm so that new training points are optimally selected to minimize this uncertainty.  As a result, in the proposed multiscale coupling framework, only a few expensive DPD simulations will be performed to provide the effective constitutive relation of the polymeric fluids to close the continuum equations. Consequently, the total computational efficiency will be significantly increased compared to traditional message-passing multiscale approaches.

The reminder of this paper is organized as follows: In Section~\ref{sec:2}, we introduce the details of macroscopic continuum equations and mesoscopic particle-based model of polymeric fluids. In Section~\ref{sec:3} we present the details of applying an active learning scheme to multiscale simulations, and demonstrate its effectiveness through numerical tests. Finally, we conclude with a brief summary and discussion in Section~\ref{sec:4}.

\section{Methods}\label{sec:2}
\subsection{Continuum Method}\label{sec:2_1}
Under the continuum hypothesis, the governing equations for an incompressible fluid in isothermal systems are the equations of continuity and momentum~\cite{2013Karniadakis},
\begin{equation}\label{eq:Gov}
\begin{split}
  \nabla\cdot \mathbf{u} &= 0, \\
  \rho\frac{\partial \mathbf{u}}{\partial t} +\rho(\mathbf{u}\cdot\nabla)\mathbf{u} = &-\nabla p+\nabla\cdot\bm{\tau},
\end{split}
\end{equation}
where $t$, $\rho$ and $p$ represent time, density and pressure, respectively; $\mathbf{u}$ is the velocity vector and $\bm{\tau}$ the stress tensors. In general, Eq.~\eqref{eq:Gov} can be non-dimensionalized to reduce the number of variables in the problem by defining characteristic scales, i.e., $L_0$, $U_0$, $\eta_0$ and $\rho_0$ for the length, velocity, viscosity and density scales, respectively. Then, all other physical variables can be scaled by these characteristic quantities, i.e., $x=x^*/L_0$, $u=u^*/U_0$, $\eta=\eta^*/\eta_0$, $\rho=\rho^*/\rho_0$, $t=t^*U_0/L_0$, and $p=p^*L_0/(\eta_0U_0)$.

To solve the macroscopic continuum equations in the form of Eq.~\eqref{eq:Gov}, a constitutive model is required to compute the stress tensor $\bm{\tau}$ in terms of various kinematic tensors, such as strain tensor and strain-rate tensor. Given a flow field, the rate of deformation tensor or strain rate tensor ${\bm \epsilon}$ can be computed by
\begin{equation}\label{eq:strain}
  {\bm \epsilon} = \frac{1}{2}\left[\nabla\mathbf{u}+(\nabla\mathbf{u})^T\right],
\end{equation}
in which $T$ denotes the transposition operator. For Newtonian fluids, the local stress is proportional to the instantaneous rate of fluid deformation, which specifies a simple constitutive equation
\begin{equation}\label{eq:newton_constitutive}
  {\bm \tau} = \eta\cdot 2{\bm \epsilon},
\end{equation}
where the constant $\eta$ is the dynamic viscosity of the Newtonian fluid related to the kinematic viscosity $\nu$ through $\eta=\rho\nu$. However, non-Newtonian fluids do not follow this linear relation between shear stress and shear strain rate given by Eq.~\eqref{eq:newton_constitutive}. Let $\tilde{\eta}$ be the apparent dynamic viscosity of a fluid defined as the ratio of shear stress to shear strain rate. Rather than a constant, the apparent viscosity $\tilde{\eta}$ of non-Newtonian fluids shows non-linear dependence on either shear rate or deformation history, wherein the shear rate dependence of $\tilde{\eta}$ can categorize non-Newtonian fluids into several types, namely pseudoplastic (shear-shinning) fluids, dilatant (shear-thickening) fluids and viscoplastic fluids~\cite{2001Balmforth,2010Chhabra}. Each of them has specified phenomenological formulas with tunable parameters to model the constitutive relation. Here, we will not specify any prior expression of constitutive relation, and compute it instead from mesoscopic simulations of microstructure dynamics of the non-Newtonian fluid.

The non-dimensionalized equation~\eqref{eq:Gov} is discretized and solved by FVM on an unstructured non-staggered grid. Specifically, the convection terms are discretized by the second-order DCQ-QUICK scheme, while Green's theorem is used to calculate the diffusive term.  The Euler method is used for time discretization of steady flows. Pressure-velocity coupling is achieved with the SIMPLE algorithm using a pressure correction~\cite{2001Ferziger}.

\subsection{Mesoscopic Method}
For a given velocity field, the stress tensor computed from micro/meso-scale dynamics simulations by ensemble averages always contains statistical errors~\cite{1989Allen}. To ensure the local ensemble average of stress tensor to be accurate, each FVM element should contain enough polymer molecules. Ideally, one should have at least $10^2-10^3$ molecules in each micro/meso-scopic system~\cite{2004Keunings}, which is usually very computational expensive for performing brute-force atomistic simulations. To this end, coarse-grained (CG) approaches drastically simplify the atomistic dynamics by eliminating fast degrees of freedom and model the unresolved details by stochastic dynamics to capture correct collective behaviors at CG levels~\cite{2013Saunders,2017Espanol,2017ZLi-DPD}. Hence, CG methods, including coarse-grained molecular dynamics (CG-MD) methods~\cite{2013Noid,2012H.Li,2012Li_Ha}, lattice-Boltzmann method~\cite{2001Succi}, multiple particle colliding dynamics~\cite{2009Gompper} and the DPD method~\cite{1992Hoogerbrugge,2017Espanol}, provide economical simulation paths to capture collective dynamics of complex fluids on larger temporal and spatial scales beyond the capability of traditional atomistic simulations. Among these CG methods, the DPD method has its roots in microscopic dynamics~\cite{2007Kinjo} and it can be derived directly from atomistic dynamics by applying the Mori-Zwanzig formalism~\cite{2010Hijon,2014ZLi_SM,2015ZLi-MZ,2017ZLi-MZ}. Moreover, DPD conserves the momentum of the system rigorously and generates correct hydrodynamic behavior of fluids at the mesoscale~\cite{1995Espanol_PRE,1997Marsh}. Therefore, in the present work, we use the DPD method to simulate the rheological dynamics of polymeric fluids at the mesoscale level.

The equation of motion of a DPD particle is governed by the Newton's second law~\cite{1997Groot},
\begin{equation}\label{eq:DPD particle}
\frac{d\mathbf{r}_{i}}{dt} =\mathbf{v}_{i},\qquad m_i\frac{d\mathbf{v}_{i}}{dt}=\mathbf{F}_i=\sum_{j\neq i}\mathbf{F}_{ij},
\end{equation}
in which $t$ denotes time and $m_i$ is the mass of the particle $i$. Also, $\mathbf{r}_i$, $\mathbf{v}_i$ and $\mathbf{F}_i$ represent position, velocity and force vectors, respectively. The pairwise interaction between two particles $i$ and $j$ consists of the conservative force $\mathbf{F}_{ij}^{C}$, dissipative force $\mathbf{F}_{ij}^{D}$£¬ and random force $\mathbf{F}_{ij}^{R}$, which are given by:
\begin{equation}\label{eq:DPD three forces}
\begin{split}
&\mathbf{F}_{ij}^{C}= a\omega_C(r_{ij})\mathbf{e}_{ij},\\
&\mathbf{F}_{ij}^{D}=-\gamma\omega_{D}\left(r_{ij}\right)\left(\mathbf{v}_{ij}\cdot\mathbf{e}_{ij}\right)\mathbf{e}_{ij},\\
&\mathbf{F}_{ij}^{R}= \sigma\omega_{R}\left(r_{ij}\right)\xi_{ij}\mathbf{e}_{ij},
\end{split}
\end{equation}
where $r_{ij}=\left|\mathbf{r}_{ij}\right|$ with $\mathbf{r}_{ij}=\mathbf{r}_{i}-\mathbf{r}_{j}$ is the distance between particle $i$ and $j$, $\mathbf{e}_{ij}=\mathbf{r}_{ij}/r_{ij}$ is the unit vector from particle $j$ to $i$, and $\mathbf{v}_{ij}=\mathbf{v}_{i}-\mathbf{v}_{j}$ is their velocity difference.
The weight functions $\omega_C(r)$, $\omega_D(r)$ and $\omega_R(r)$ are defined with a cutoff radius $r_c$ beyond which these weight functions vanish. The coefficients $\alpha$, $\gamma$ and $\sigma$ determine the strength of each force.
$\xi_{ij}=\xi_{ji}$ is a Gaussian white noise $\langle\xi_{ij}(t)\rangle=0$ and $\langle \xi_{ij}(t)\xi_{kl}(t')\rangle=(\delta_{ik}\delta_{jl}+\delta_{il}\delta_{jk})\delta(t-t')$, in which $\delta_{ij}$ is the Kronecker delta and $\delta(t-t')$ is the Dirac delta function~\cite{1995Espanol_EPL}.
The dissipative force $\mathbf{F}^D_{ij}$ and random force $\mathbf{F}^R_{ij}$ are related via the fluctuation-dissipation theorem by satisfying
\begin{equation}\label{eq:FDT}
\sigma^{2}=2\gamma k_{B}T, \qquad \omega_D(r)=\omega_R^2(r),
\end{equation}
where $k_B$ is the Boltzmann constant and $T$ the temperature. A typical choice for the weight functions is $\omega_C(r)=1-r/r_c$ and $\omega_D(r)=\omega_R^2(r)=(1-r/r_c)^2$ with $r_c$ being the cutoff radius.

For explicitly modeling the dynamics of polymer chains, we employ the bead-spring model where the polymer is modeled as a chain of $N$ beads connected by a finitely extensible nonlinear elastic (FENE) spring~\cite{1990Kremer}. In particular, the FENE potential is the form~\cite{2010Fedosov}
\begin{equation}\label{eq:FENE}
U_{\rm FENE}=-\frac{k_{s}R^2}{2}\log\left[1-\left(\frac{r_{ij}}{R}\right)^2 \right],
\end{equation}
where $k_{s}$ is the spring constant and $R$ the maximum bond extension of the FENE spring. In DPD simulations, the FENE bond is short and stiff enough for avoiding unphysical bond-crossings~\cite{2014ZLi_SM}, and hence the correct polymeric rheology with chain entanglements can be simulated.

\section{Numerical Implementation and Results}\label{sec:3}
\begin{figure}[h!]
\centering
\includegraphics[width=0.65\textwidth]{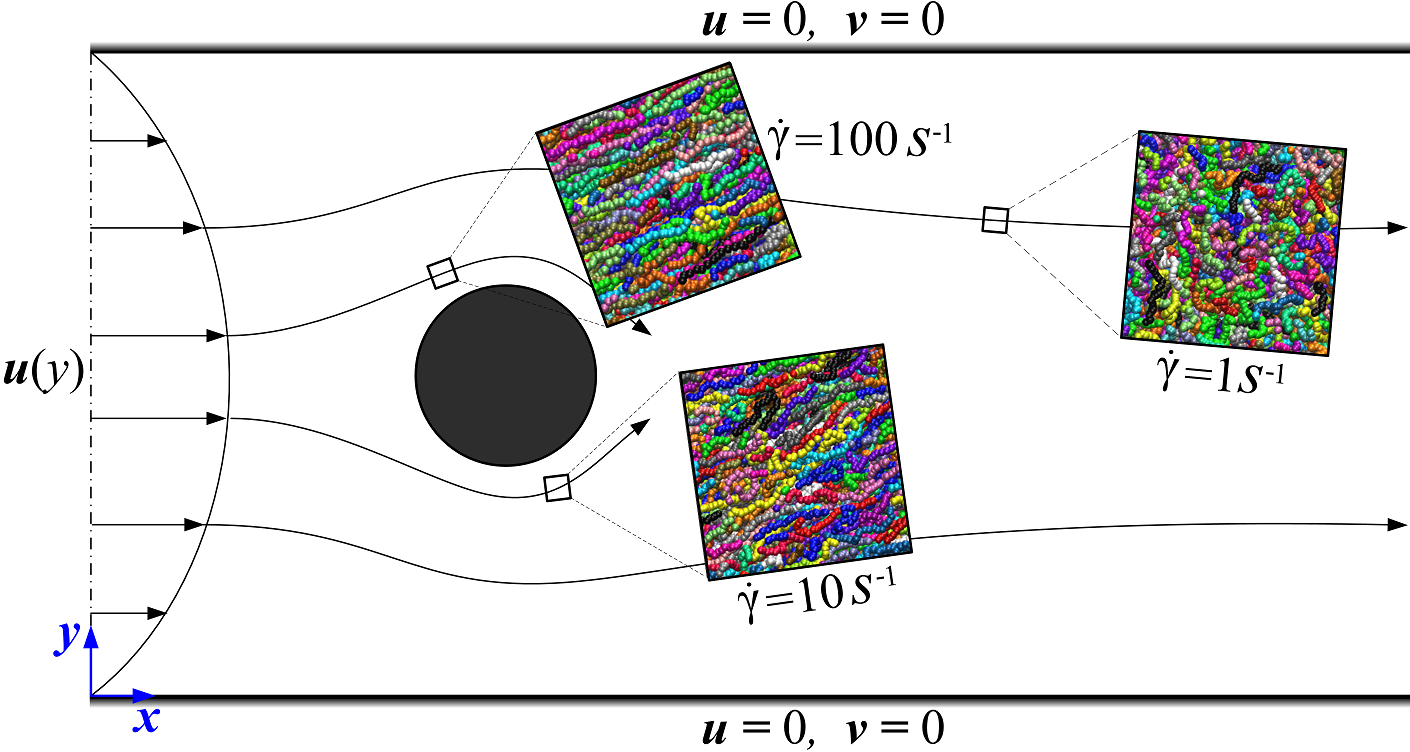}
\caption{Schematic of the proposed multiscale modeling of non-Newtonian fluids and the setup of boundary conditions for the continuum method. The continuum solver passes local flow fields to mesoscopic simulations, while the constitutive relation of the non-Newtonian fluid is actively learned from adaptively running selected mesoscopic dynamics. The three snapshots of polymer chains shown illustrate that a polymeric fluid can exhibit significantly different underlying microstructures of polymer chains at different shear rates.}\label{FIG 1}
\end{figure}

A non-Newtonian polymeric fluid flowing past a cylinder between two parallel plates, as shown in Fig.~\ref{FIG 1}, is used as a benchmark system to test the effectiveness of applying the active learning scheme to multiscale simulations. At the macroscopic level, the polymeric fluid is modeled by the continuum equations in the form of Eq.~\eqref{eq:Gov}, which is discretized and solved by FVM as described in Sec.~\ref{sec:2_1}. More specifically, we perform the simulation in a two-dimensional rectangular computational region of $60L_0\times30L_0$ inside which a static circular cylinder with diameter of $L_0=2~cm$ is placed at $(x,y)=(15L_0, 15L_0)$, as shown in Fig.~\ref{FIG 1}. The coordinates $x$ and $y$ represent the stream-wise and crossflow directions, and the corresponding velocity components are denoted by $u$ and $v$, respectively. The computational domain is discretized by $518\,614$ triangular elements with the smallest mesh size being $0.01L_0$ near the cylinder surface.

A velocity profile of the plane Poiseuille flow is imposed at the inlet boundary,
\begin{equation}
u(y)=\frac{4U_0}{L_y^2}\cdot y(L_y-y), \qquad v(y)=0,
\end{equation}
where $L_y=30L_0$ is the channel width and $U_0$ is the maximum inlet velocity. A Neumann boundary condition for velocity ($\partial u/\partial x = \partial v/\partial x = 0$) and a fixed value for pressure $p=p_0$ are applied at the outlet. The surfaces of the upper and lower walls and also the circular cylinder are set to be solid wall with a no-slip ($u=0,~v=0$) boundary condition.

The fluid has a zero shear viscosity of $\nu_0=1.0\times10^{-4}~m^2/s$. The Reynolds number is defined by $Re=U_0L_0/\nu_0$, where the characteristic length $L_0$ is the diameter of the cylinder. Flows at different $Re$ can be generated by applying various inlet velocities, i.e., $U_0=5~cm/s$ corresponds to a flow at $Re=10$. At macroscale level, the non-Newtonian behavior generally falls into three categories~\cite{2010Chhabra} -- (i) shear stress depends only on the current shear rate, (ii) shear stress depends on the duration of shearing as well as the previous kinematic history, and (iii) materials exhibit viscoelastic behavior with ability to store and recover shear energy. The first kind of non-Newtonian fluids are called generalized Newtonian fluid, which is the type we will focus on in this work. Let $\epsilon_{xx}=\partial u/\partial x$, $\epsilon_{yy}=\partial v/\partial y$ and $\epsilon_{xy}=\epsilon_{yx}=(\partial u/\partial y + \partial v/\partial x)/2$ be the components of the strain rate tensor $\bm \epsilon$. By implementing a tensor transformation, we can obtain the maximum shear strain rate given by~\cite{1993Boresi}
\begin{equation}\label{eq:max shear strain}
  \dot{\gamma}=\sqrt{\frac{1}{4}(\epsilon_{xx}-\epsilon_{yy})^2 + \epsilon_{xy}^2},
\end{equation}
which is taken as the local shear rate on each FVM element to compute the local stress tensor. In addition to the maximum shear strain rate, $\dot{\gamma}$ can be also defined by the second invariant of the strain rate tensor~\cite{2011Chai,1994Macosko}, i.e., $\dot{\gamma}=\sqrt{2~{\rm tr}(\bm \epsilon^2)}$. Then, the global minimum and the global maximum of $\dot{\gamma}$ obtained from the FVM transient solution will indicate (on-the-fly) the range of effective constitutive relation required to close the continuum equations.

A polymeric fluid can exhibit significantly different underlying microstructures of polymer chains at different shear rates, as illustrated by the three snapshots of polymer chains in Fig.~\ref{FIG 1}. At low shear rates, the configuration of polymer chains can be characterized by a random coil, which results in a large flow resistance. However, the polymer chains trend to align in the direction of the flow and generate a decreasing flow resistance as the shear rate increases. As a results, unlike Newtonian fluids whose shear stress is a linear function of strain rate, the macroscopic rheology of polymeric fluids shows non-linear dependence on the strain rates. Using multiscale simulations, we compute the nonlinear constitutive relation of polymeric fluids directly from DPD simulations rather than using any empirical expression for the constitutive relation.

\begin{figure}[b!]
\centering
\includegraphics[width=0.6\textwidth]{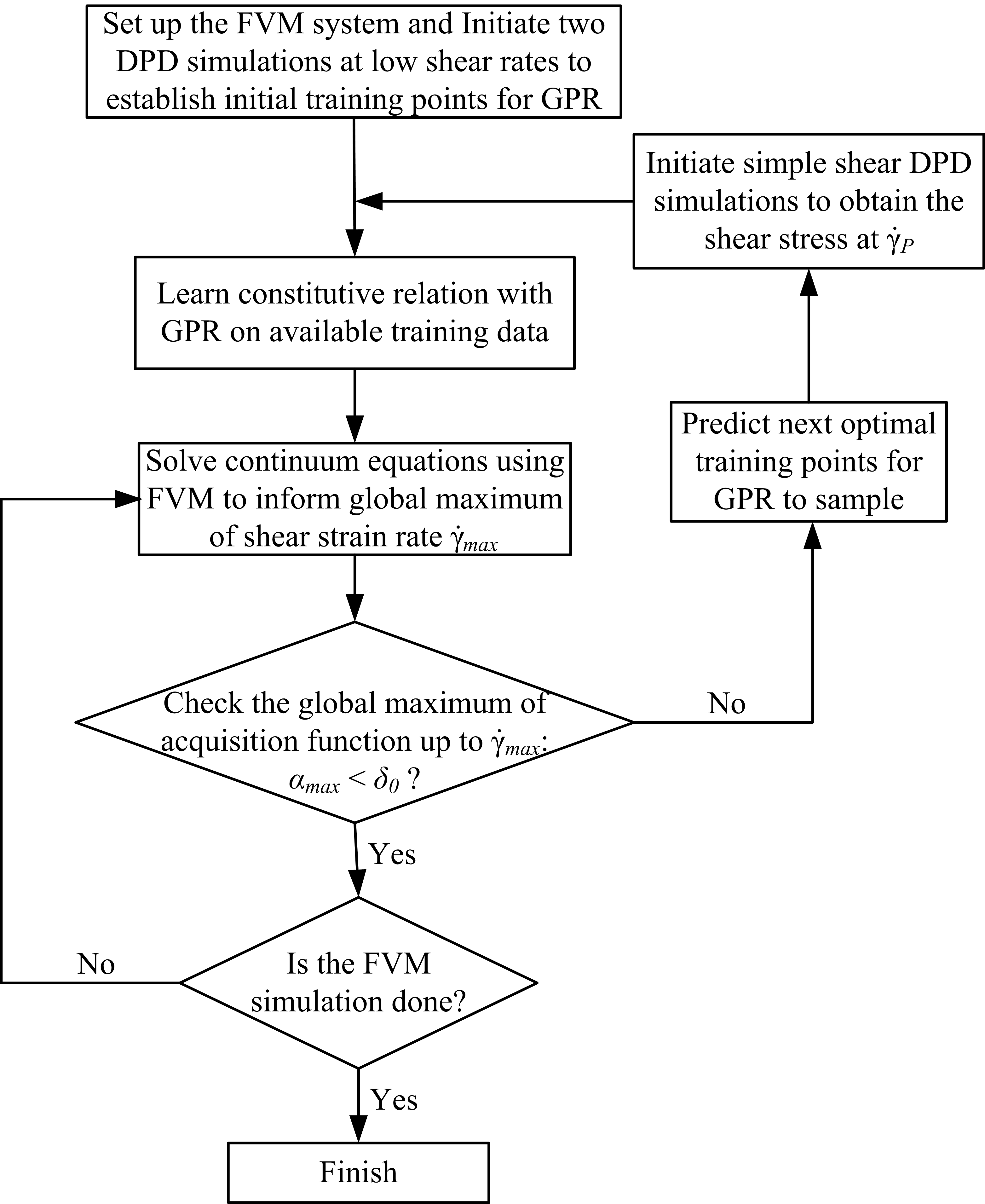}
\caption{Workflow of applying an active learning scheme to multiscale modeling of non-Newtonian fluids, where the constitutive relation is learned on-the-fly  from mesoscale DPD simulations as the effective model closure used in the macroscale FVM solver.}
\label{fig:workflow}
\end{figure}

At the mesoscopic level, the polymeric fluid is modeled by DPD with explicit polymer chains. A DPD system is constructed with $3\,750$ polymer chains in a computational box of $50\times25\times25$ in reduced DPD units. Each chain consists of $25$ DPD beads connected by the FENE spring given by Eq.~\eqref{eq:FENE} with $k_s=50$ and $R=1.0$. The non-bonded interactions between DPD particles are computed by Eq.~\eqref{eq:DPD three forces} with a parameter set $a=25, \gamma=4.5, \sigma=3, r_c=2.0$. A simple shear DPD simulation at a selected shear rate $\dot{\gamma}_p$ will be performed to obtain the shear stress $\tau$ in the polymeric fluid at $\dot{\gamma}_p$, wherein we compute the stress tensor by the Irving-Kirkwood formula~\cite{1950Irving,2012Yang}. To map physical units in macroscopic equations to reduced DPD units, the DPD system is constructed with a length unit $[L]=5.0\times 10^{-6}~m$, a time unit $[T]=2.63\times 10^{-4}~s$, and a mass unit $[M]=1.25\rho_n^{-1}\times 10^{-13}~kg$ with $\rho_n=3.0$ being the number density of DPD particles. Then, all the physical quantities with physical units can be mapped to reduced DPD units~\cite{2017ZLi-DPD} using the three basic units $[L]$, $[T]$ and $[M]$, i.e., the shear rate is converted by the shear rate unit $[\dot{\gamma}]=1/[T]=3.80\times 10^3~s^{-1}$, the shear stress is converted by the stress unit $[\tau]=[M][L]^{-1}[T]^{-2}= 0.12~Pa$, and the kinematic viscosity is converted by $[\nu]=[L]^2/[T]=9.5\times 10^{-8}~m^2/s$.

In the multiscale simulation, the macroscopic continuum equations are discretized and solved by FVM on unstructured non-staggered grid, and the mesoscopic dynamics is simulated using the DPD method. Figure~\ref{fig:workflow} shows the workflow of applying an active learning scheme to multiscale modeling of non-Newtonian fluids, where the effective constitutive relation is  constructed on-the-fly to close the macroscopic momentum equation in FVM. In particular, we first set up the FVM system and perform two DPD simulations of simple shear flow at low shear rates close to the zero-shear rate limit to establish the initial training points for GPR. Then, the GPR-informed constitutive relation can be used to calculate the shear stress on each FVM element in terms of local stain rate tensor. In practice, we start with a FVM system of a flow past a cylinder at $Re=10$. Two DPD simulations of the polymeric fluid in simple shear flows at shear rates $\dot\gamma=1.0~s^{-1}$ ($2.63\times 10^{-4}$ in reduced units) and $2.0~s^{-1}$ ($5.26\times 10^{-4}$ in reduced units) are performed to generate the initial training data for GPR. The GPR prediction estimates not only the mean of the apparent viscosity $\nu(\dot\gamma)$, but also the uncertainty, i.e., the standard deviation of GPR prediction $\sigma(\dot\gamma)$. Since the range of shear rates in FVM changes by several orders of magnitude, we perform the GPR on the log-space, e.g., $\log\dot\gamma$ versus $\log\nu$. To obtain the next optimal sampling point, we define an acquisition function as $\alpha=\sigma(\dot\gamma)$ and choose an acceptance criterion $\max(\alpha)\le\delta_{\rm tol}$ for the GPR-informed constitutive relation, in which $\delta_{\rm tol}=0.01$ is a predefined tolerance.

\begin{figure}[b!]
\centering
\includegraphics[width=0.6\textwidth]{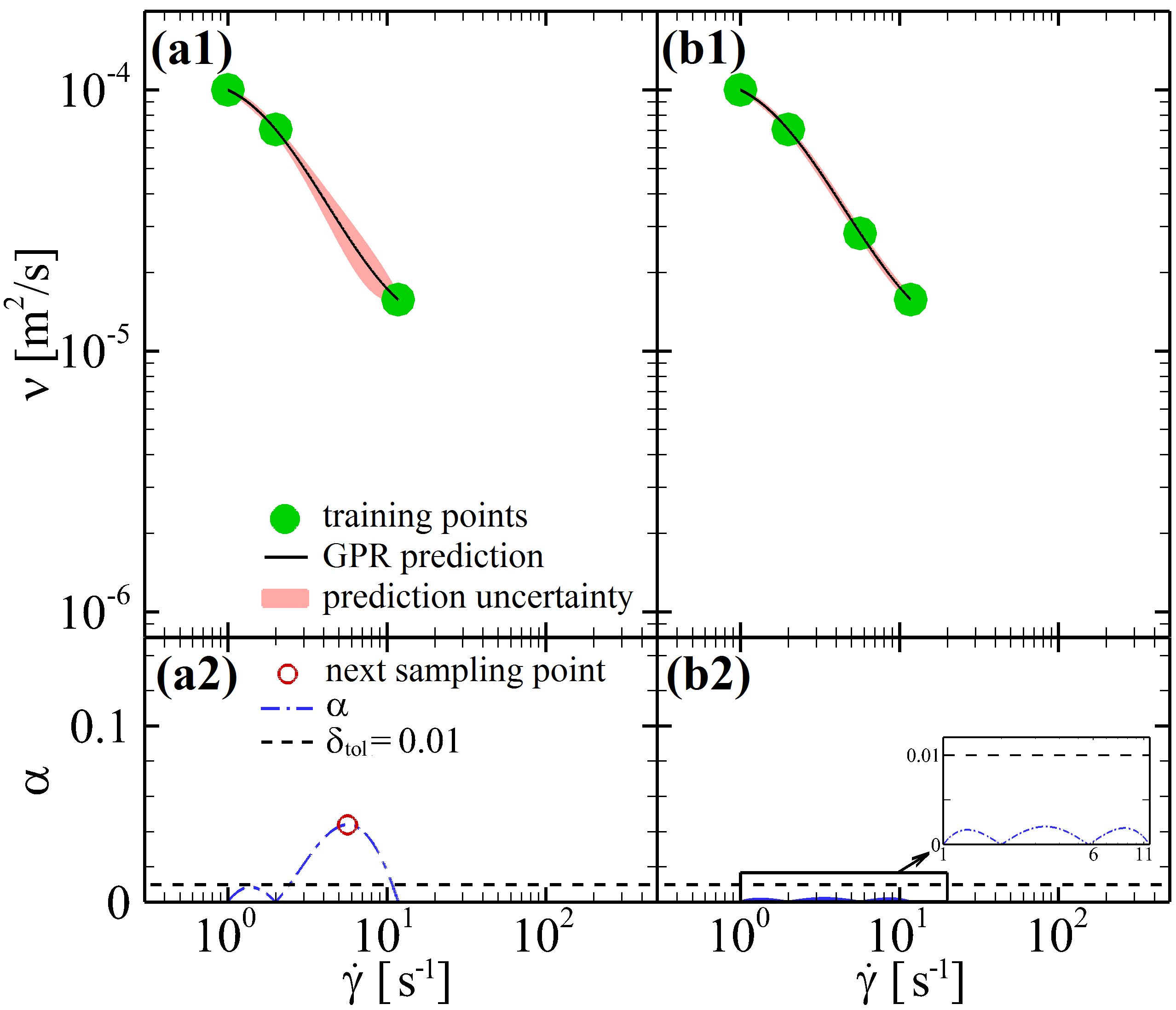}
\caption{
GPR-informed constitutive relations based on (a1) three training points and (b1) four training points obtained from DPD simulations. The filled circles represent training data, and solid lines are GPR prediction with prediction uncertainties visualized by the $95\%$ confidence interval shown as the shaded area. The dash-dotted line in (a2) shows the magnitude of the acquisition function $\alpha(\dot\gamma)$, whose maximum indicates the next sampling point $\dot\gamma_{p2}=5.67~s^{-1}$ for interpolation.}
\label{FIG:First_int}
\end{figure}

With the constitutive relation constructed based on the initial training points, we can then advance the FVM system. After a single time step integration, the transient FVM flow field has a global maximum of $\dot\gamma_{\rm max}$ being $11.78~s^{-1}$, at which the magnitude of $\alpha$ reaches its maximum value far beyond the designed tolerance $\delta_{\rm tol}$. This breaks our acceptance criterion and we need additional DPD runs to reduce the magnitude of $\alpha$. Subsequently, we take the shear rate $\dot\gamma_{p1}$, where $\alpha$ reaches its maximum, as the next sampling point, and hence the FVM solver starts a request to perform a simple shear DPD simulation at $\dot\gamma_{p1}=11.78~s^{-1}$. Figure~\ref{FIG:First_int}(a1) shows the GPR-informed constitutive relation based on three DPD simulations, in which the filled circles represent the training points, the solid line denotes the mean values of the GPR prediction, and the prediction uncertainties of GPR-informed constitutive relation are visualized by the $95\%$ confidence interval shown as the shaded area. Figure~\ref{FIG:First_int}(a2) plots the magnitude of corresponding $\alpha(\dot\gamma)$ up to $\dot\gamma_{\rm max}=11.78~s^{-1}$, where the global maximum of $\alpha(\dot\gamma)$ clearly indicates the next optimal sampling point $\dot\gamma_{p2}=5.67~s^{-1}$ for interpolation. With one additional data point, we have a better GPR-informed constitutive relation displayed in Fig.~\ref{FIG:First_int}(b1), in which the magnitude of  $\alpha(\dot\gamma)$ is reduced below the tolerance $\delta_{\rm tol}$, as shown in Fig.~\ref{FIG:First_int}(b2). This process, which is described by the right loop in the workflow of Fig.~\ref{fig:workflow}, can be repeated until the prediction uncertainty of GPR-informed constitutive relation is smaller than a prescribed tolerance $\delta_{\rm tol}$ over the range of shear rates up to $\dot\gamma_{\rm max}$.

\begin{figure}[b!]
\centering
\includegraphics[width=0.6\textwidth]{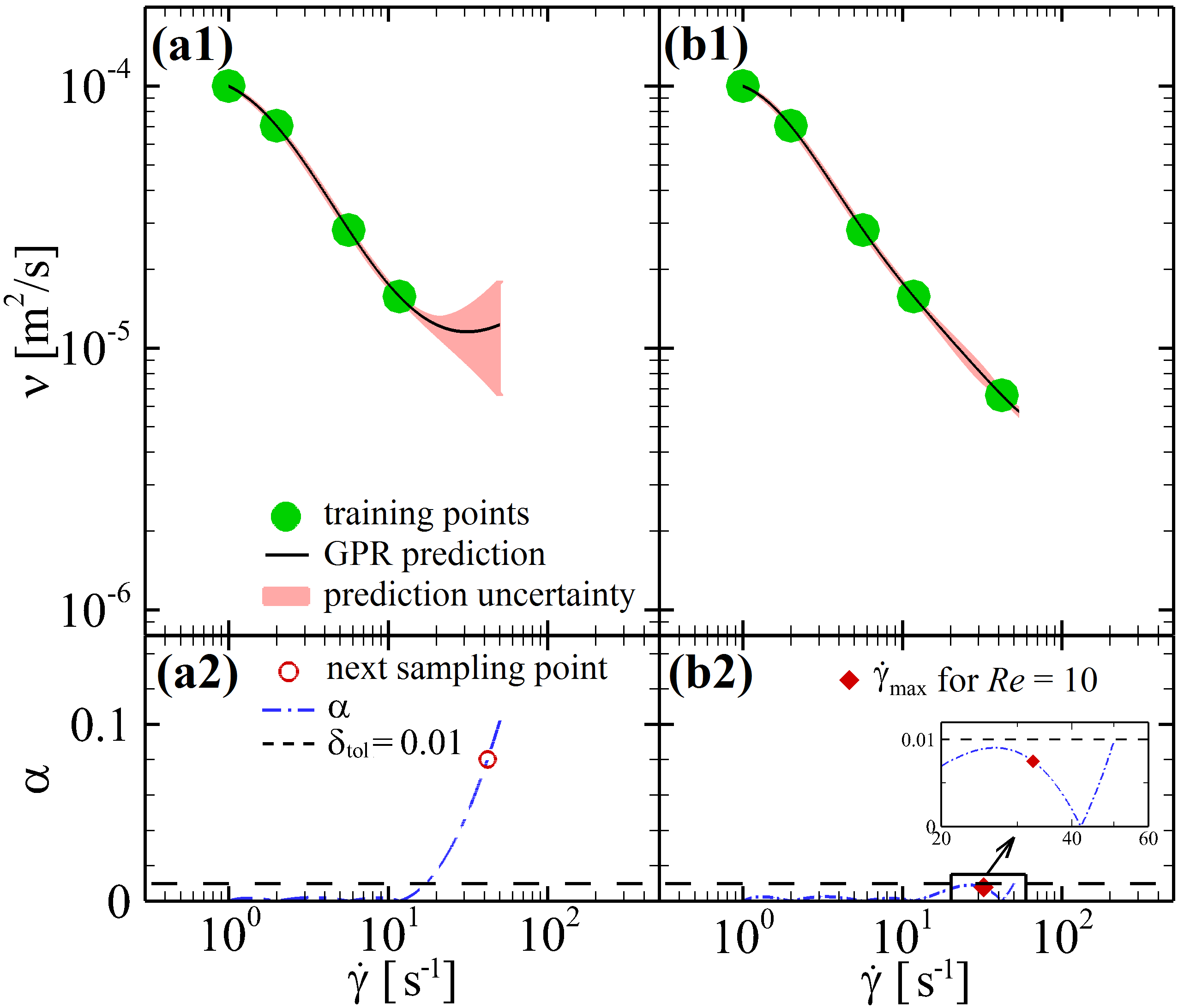}
\caption{GPR-informed constitutive relations based on (a1) four training points and (b1) five training points obtained from DPD simulations. The dash-dotted line in (a2) shows the magnitude of the acquisition function $\alpha(\dot\gamma)$. The next sampling point $\dot\gamma_{p3}=41.9~s^{-1}$ for extrapolation is determined by $\alpha(\dot\gamma_p)=\beta\delta_{\rm tol}$, where $\beta=8$ is used in this case. The filled diamond in (b2) shows the global maximum shear rate $\dot\gamma_{\rm max}=32.5~s^{-1}$ in the flow of $Re=10$, which indicates that the range of shear stain rate covered by the GPR-informed constitutive relation shown in (b1) (up to $\dot\gamma=50.2~s^{-1}$) is broader than the current requirement for the flow at $Re=10$.}
\label{FIG:First_ext}
\end{figure}

Since the FVM system is initialized with a zero-velocity flow field, the global maximum of strain rate $\dot{\gamma}_{\max}$ in the FVM system changes with time before it reaches the steady state. Therefore, we need to monitor the magnitude of the acquisition function $\alpha(\dot\gamma)$ (up to $\dot{\gamma}_{\max}$) to on-the-fly inform the necessity of performing additional DPD simulations. As shown in Fig.~\ref{FIG:First_ext}(a1), the prediction uncertainty increases sharply with the shear strain rate. Let $\dot\gamma_{\rm gpr}$ be the maximum strain rate covered by a valid GPR-informed constitutive relation with $\alpha<\delta_{\rm tol}$. Then, we may need additional sampling points for extrapolation once the global maximum of strain rate $\dot\gamma_{\rm max}$ in the FVM flow field becomes larger than $\dot\gamma_{\rm gpr}$. In general, including more training points can significant reduce the magnitude of $\alpha(\dot\gamma)$ near new added points and beyond. The next sampling point $\dot\gamma_p$ for extrapolation is determined by $\alpha(\dot\gamma_p)=\beta\delta_{\rm tol}$, where $\beta$ is a tunable parameter defining how aggressive for extrapolation. Small $\beta$ may be too safe to reduce the number of training points, but large $\beta$ may be too aggressive and requires extra interpolating points after adding the extrapolating point to satisfy the acceptance criterion. A proper value $\beta>1.0$ can be estimated by applying a Bayesian optimization based on a few attempted extrapolations. In the present study, we use $\beta=8$ indicating the next sampling point $\dot\gamma_{p3}=41.9~s^{-1}$. After adding an extrapolating point, we need to recheck the magnitude of $\alpha(\dot\gamma)$ to determine if extra interpolating points are required, as we did in Fig.~\ref{FIG:First_int}. The plot of $\alpha(\dot\gamma)$ in Fig.~\ref{FIG:First_ext}(a2) indicates the acceptance criterion $\alpha<\delta_{\rm tol}$ is satisfied and no interpolating point is needed.

\begin{figure}[b!]
\centering
\includegraphics[width=0.6\textwidth]{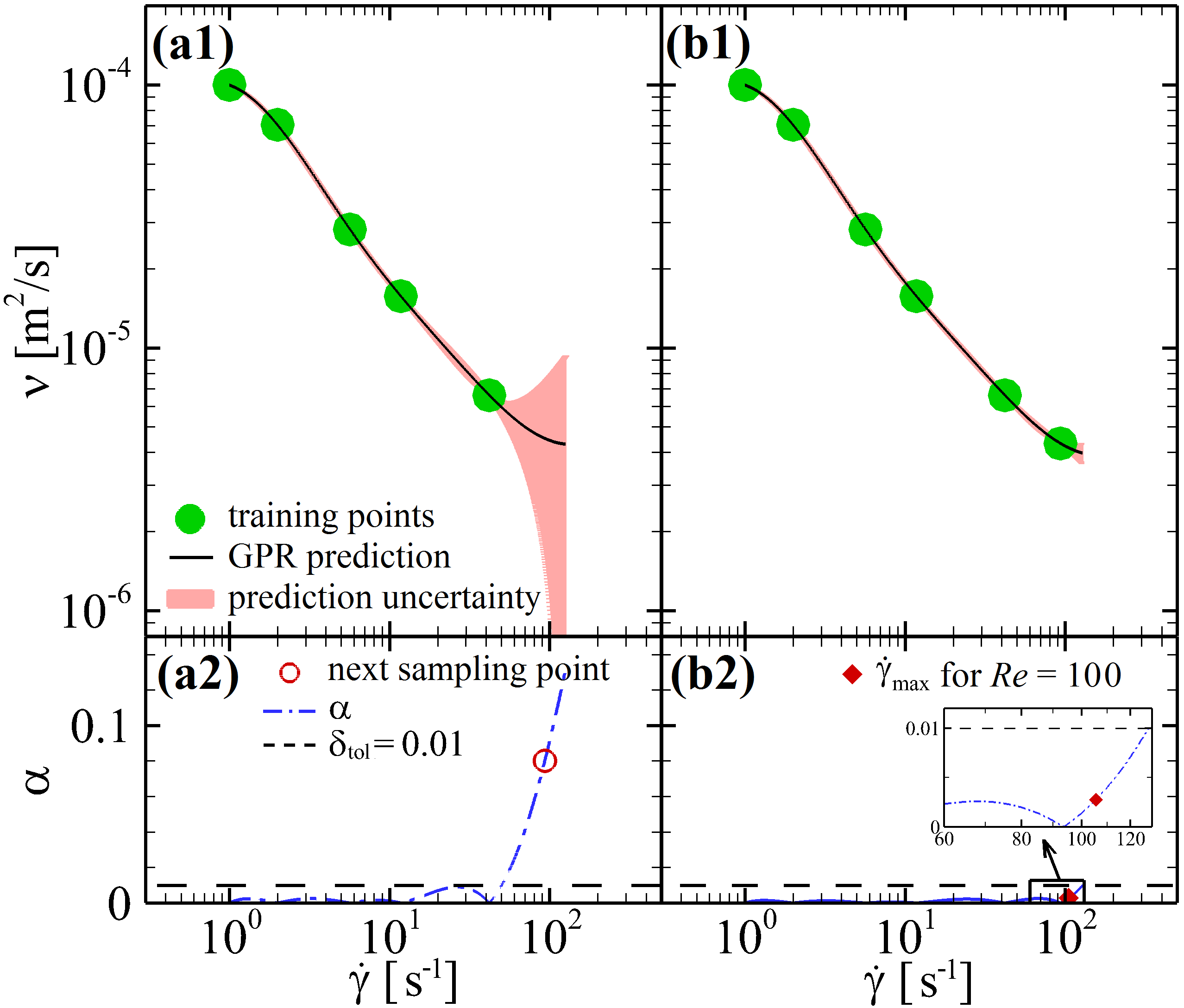}
\caption{GPR-informed constitutive relations based on (a1) five training points and (b1) six training points obtained from DPD simulations. The dash-dotted line in (a2) shows the magnitude of the acquisition function $\alpha(\dot\gamma)$. The next sampling point $\dot\gamma_{p4}=93.4~s^{-1}$ for extrapolation is determined by $\alpha(\dot\gamma_p)=\beta\delta_{\rm tol}$, where $\beta=8$ is used in this case. The filled diamond in (b2) shows the global maximum shear rate $\dot\gamma_{\rm max}=105.5~s^{-1}$ in the flow at $Re=100$. }\label{FIG:Second_ext}
\end{figure}

Figure~\ref{FIG:First_ext}(b1) illustrates the GPR-informed constitutive relation after including one additional training point at $41.9~s^{-1}$. Given the acceptance criterion ${\rm max}(\alpha)<\delta_{\rm tol}$, the plot of $\alpha(\dot\gamma)$ in Fig.~\ref{FIG:First_ext}(b2) shows that the new constitutive relation is valid up to $\dot\gamma=50.2~s^{-1}$. The filled diamond in Fig.~\ref{FIG:First_ext}(b2) at $\dot\gamma_{\rm max}=32.5~s^{-1}$ is the global maximum strain rate in the flow at $Re=10$. It indicates that the GPR-informed constitutive relation in Fig.~\ref{FIG:First_ext}(b1), which is based on five training points obtained from DPD simulations, is broader than the current requirement for the flow at $Re=10$. Therefore, when the Reynolds number of the flow is increased to $Re=20$ and $30$, the GPR-informed constitutive relation in Fig.~\ref{FIG:First_ext}(b1) without any additional DPD runs can still close the continuum equations in the FVM system. If we increase the Reynolds number to $Re=100$, the maximum strain rate can go beyond the point $\dot\gamma_{\rm gpr}=50.2~s^{-1}$ in Fig.~\ref{FIG:First_ext}(b1). Then, we need to perform additional DPD simulations to include new data points. Using the same method for extrapolation in Fig.~\ref{FIG:First_ext} with $\beta=8$, as shown in Fig.~\ref{FIG:Second_ext}(a2), we carry out a DPD simulation of a sample shear flow at $\dot\gamma_{p4}=93.4~s^{-1}$ and we have a GPR-informed constitutive relation valid up to $\dot\gamma_{\rm gpr}=128.2~s^{-1}$, which is sufficient to close the FVM system for the flow of $Re=100$ whose maximum strain rate is $\dot\gamma_{\rm max}=105.5~s^{-1}$.

\begin{figure}[t!]
\centering
\includegraphics[width=0.8\textwidth]{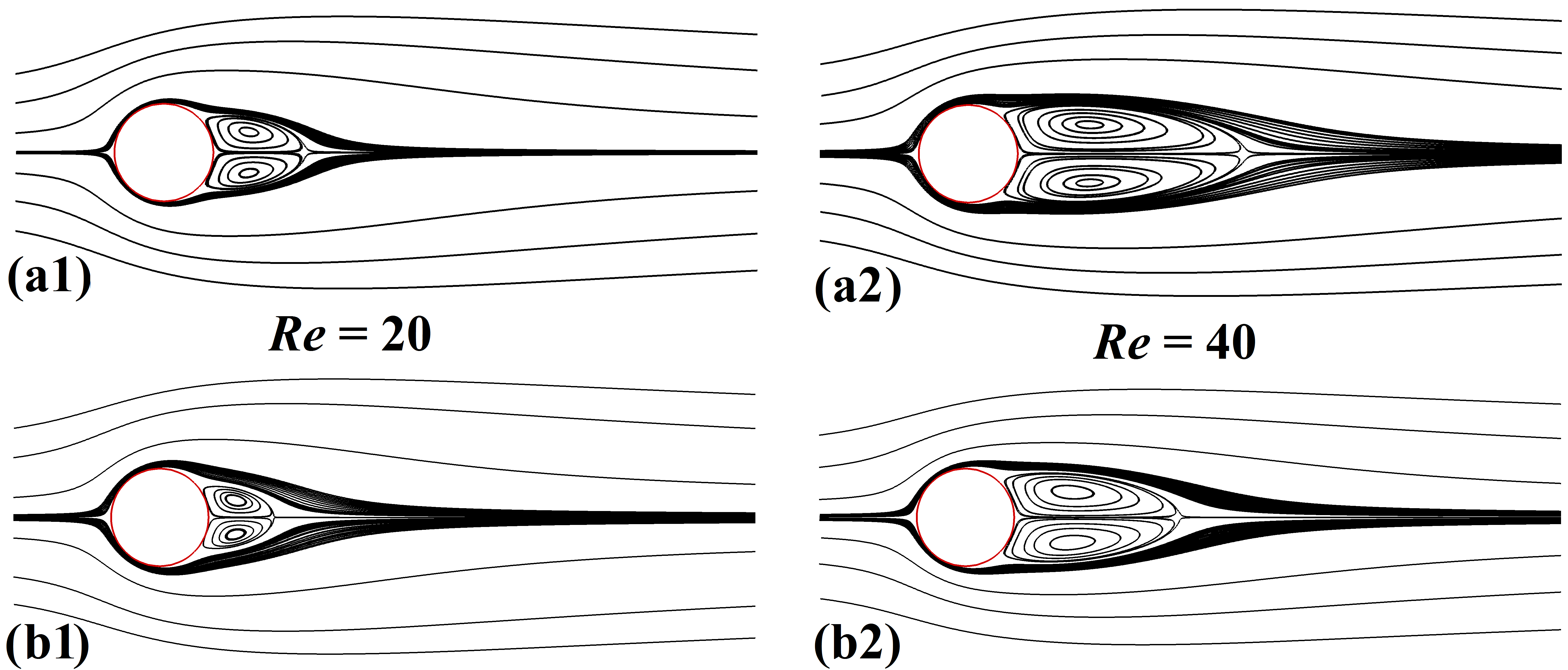}
\caption{Comparison of the streamline contours for the Newtonian fluid (a1, a2) and the non-Newtonian polymeric fluid (b1, b2) at two Reynolds numbers. The left column is for the wake flow at $Re=20$, and the right column is for $Re=40$.}\label{FIG:Streamline}
\end{figure}

Figure~\ref{FIG:Streamline} shows the streamline contours for the polymeric fluid flow over a cylinder at Reynolds numbers of $20$ and $40$, which are compared with corresponding Newtonian flows. These results show that the polymeric fluid whose constitutive relation closure is actively learned from DPD simulations has different rheological behavior from a Newtonian fluid. The wakes behind the cylinder of polymeric fluid are smaller than these of a Newtonian fluid at the same Reynolds number, which corresponds to the typical shear-thinning behavior~\cite{2012lashgari}. It is worth noting that no prior shear-thinning model is imposed in the multiscale simulation, as the non-Newtonian observation is captured by the effective constitutive relation computed from DPD simulations by assuming that the stress is determined by the viscosity in steady shear flow.

\begin{figure}[t!]
\centering
\includegraphics[width=0.5\textwidth]{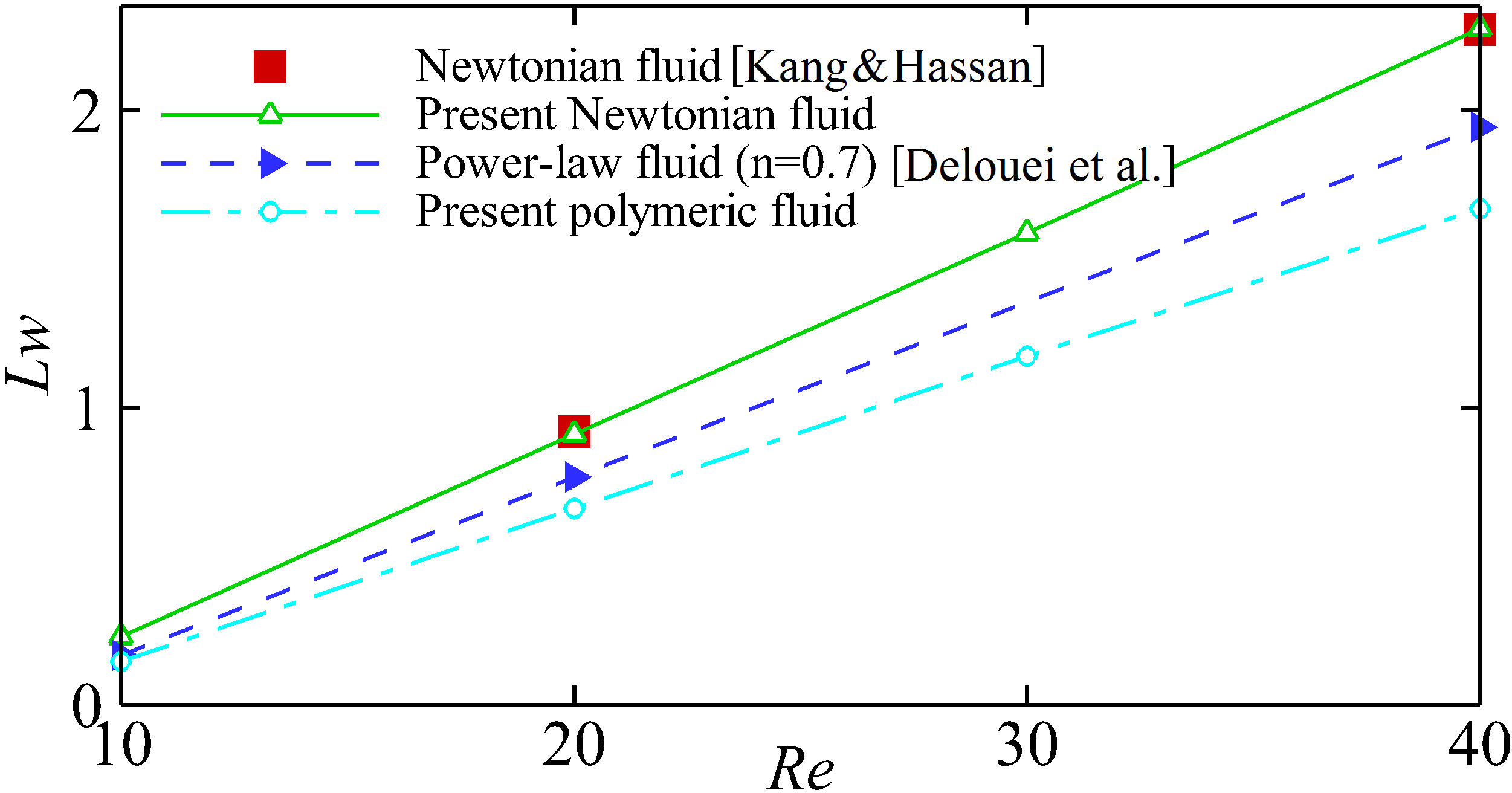}
\caption{Dependence of the recirculation length $L_{\rm w}$ on Reynolds number $Re$ for Newtonian and non-Newtonian fluids, and comparison with Kang \& Hassan's work~\cite{2011Kang} for a Newtonian fluid and Delouei et al.'s work~\cite{2014Delouei} for a power-law ($n=0.7$) fluid.}
\label{FIG:Lw}
\end{figure}

The recirculation length $L_{\rm w}$ can be used to quantify the size of a steady separation bubble behind the cylinder in Newtonian and non-Newtonian flows. $L_{\rm w}$ in Figs.~\ref{FIG:Streamline}(a1),~\ref{FIG:Streamline}(a2),~\ref{FIG:Streamline}(b1) and~\ref{FIG:Streamline}(b2) are 0.91, 2.27, 0.66 and 1.67, respectively. This indicates that the wakes of both Newtonian and polymeric fluids tend to be longer as $Re$ increases. We plot the recirculation length $L_{\rm w}$ at various Reynolds numbers in Fig.~\ref{FIG:Lw}. For Newtonian fluid flows, our results are in good agreement with a previous study of Newtonian steady flow over a circular cylinder~\cite{2011Kang}. This also validates our in-house developed FVM solver. It is shown in Fig.~\ref{FIG:Lw} that the wake length in the polymeric fluid is shorter than that in a Newtonian fluid at the same $Re$, which is similar to the behavior of a power-law shear-thinning fluid~\cite{2012lashgari,2014Delouei}.

\begin{figure}[t!]
\centering
\includegraphics[width=0.9\textwidth]{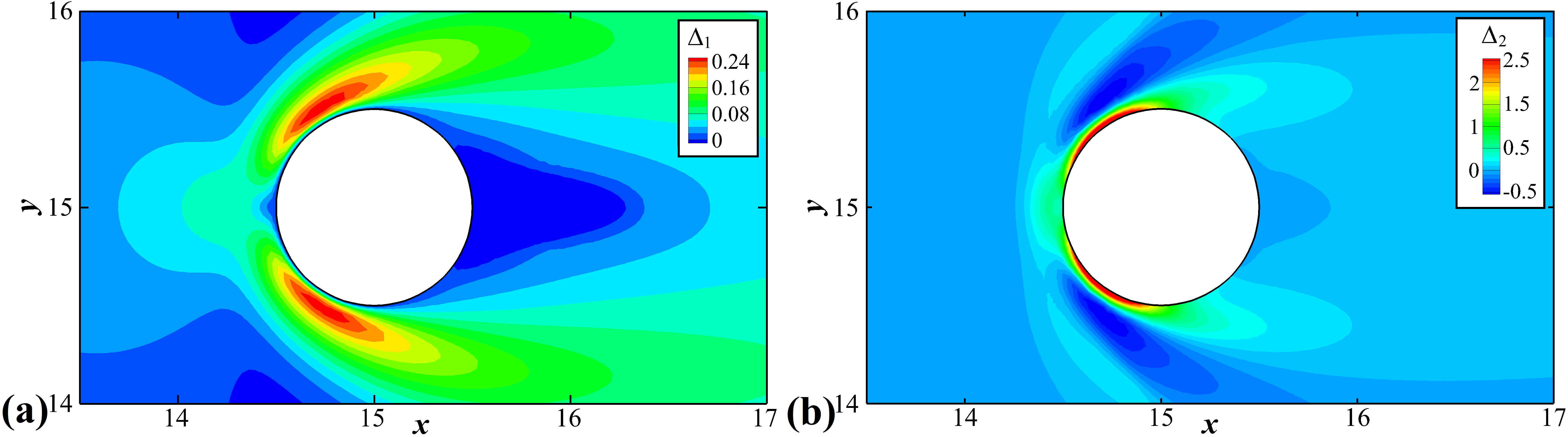}
\caption{Velocity difference $\Delta_1=u-u'$ and stain rate difference $\Delta_2=\dot\gamma-\dot\gamma'$ between the non-Newtonian fluid and the Newtonian fluid flowing around a cylinder at $Re=20$. Here $u$ and $\dot\gamma$ represent the velocity and the strain rate in the flow of the polymeric fluid, and $u'$ and $\dot\gamma'$ are for the flow of the Newtonian fluid.}\label{FIG:Delta}
\end{figure}

Figure~\ref{FIG:Delta} shows the velocity difference $\Delta_1=u-u'$ as well as the stain rate difference $\Delta_2={\bm \dot\gamma}-{\bm \dot\gamma}'$ between the flow field of the non-Newtonian polymeric fluid and the Newtonian fluid at $Re=20$. As shown in Fig.~\ref{FIG:Delta}(a), the flow of the polymeric fluid in the vicinity of the cylinder is accelerated compared to that of a Newtonian fluid at the same Reynolds number. It is observed in Fig.~\ref{FIG:Delta}(b) that the stain rate is increased in a layer attached to the surface of the cylinder, but it is decreased in the region away from the cylinder. For a shear-thinning fluid, the increased strain rate around the cylinder will yield a significant reduction of local viscosity of the fluid and generate a thin layer of low viscous fluid encapsulating the cylinder. However, a reduction in the stain rate in the region away from the cylinder yields high viscous fluid, which acts as an effective confinement suppressing the tendency for flow separation. Therefore, the flow separation around the cylinder is somewhat delayed in the shear-thinning polymeric fluid compared to the Newtonian fluid. As a consequence, we find in Fig.~\ref{FIG:Lw} that the recirculation length $L_{\rm w}$ of the polymeric fluid is smaller than that of the Newtonian fluid at the same Reynolds number.

\begin{figure}[t!]
\centering
\includegraphics[width=0.55\textwidth]{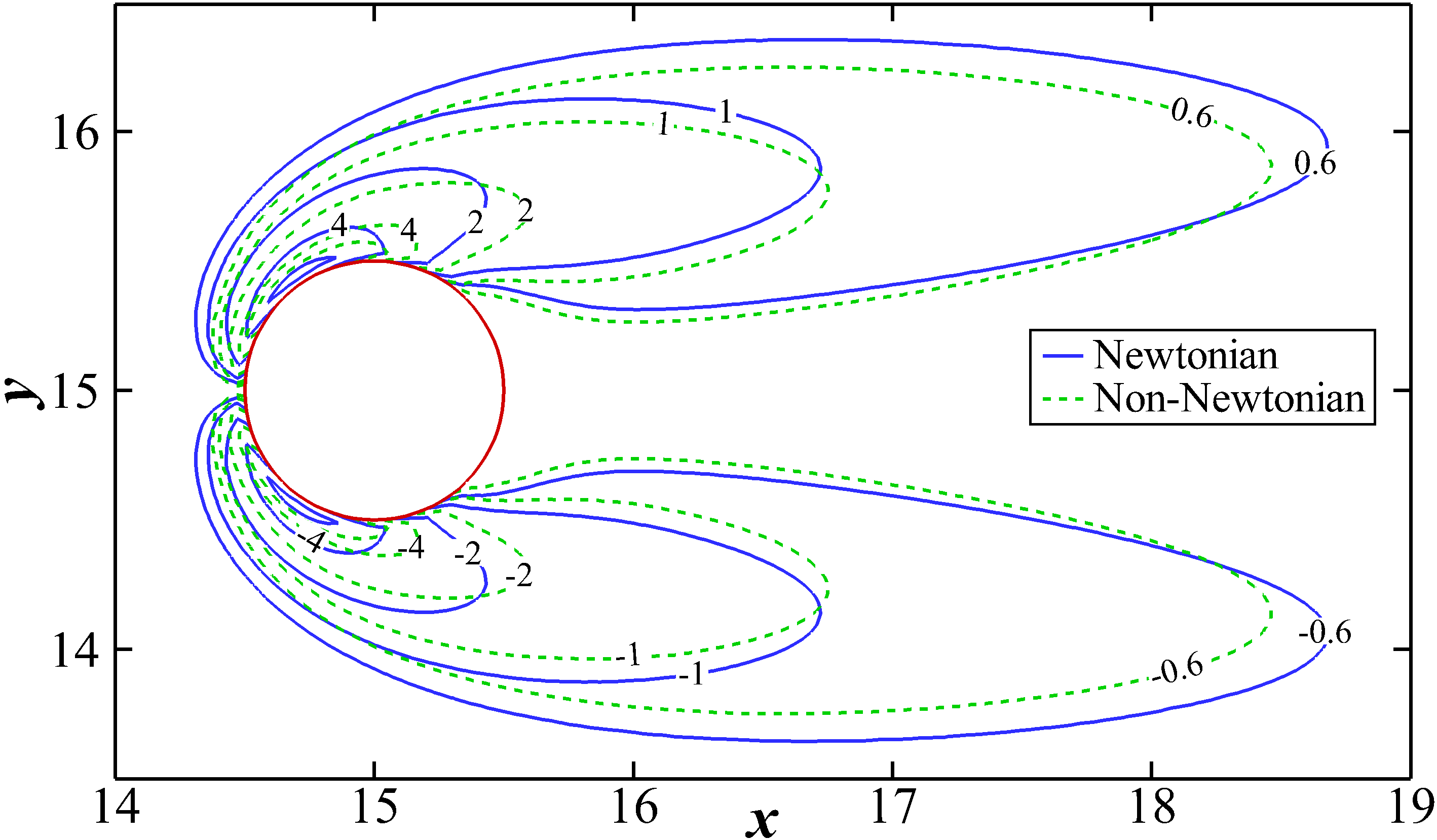}
\caption{Comparison of contour lines of the vorticity between the wake flow around a cylinder at $Re=20$ for the Newtonian fluid and the non-Newtonian polymeric fluid whose constitutive relation closure is computed from DPD simulations.}
\label{FIG:Vorticity}
\end{figure}

The difference between the wake flow of the Newtonian fluid and the non-Newtonian fluid can also be visualized by the contour plot of the vorticity $\omega=\partial v/\partial x - \partial u/\partial y$. Figure~\ref{FIG:Vorticity} shows several contour lines of vorticity for the wake flow at $Re=20$, where the solid contour lines are for the Newtonian flow, and the dashed lines for the flow of the polymeric fluid whose constitutive relation closure is computed from DPD simulations. The difference between the vorticity contours verifies that the separation around the cylinder is delayed and a slightly shrunken wake behind the cylinder is observed in the shear-thinning polymeric fluid, which is consistent with the analyses and conclusions obtained from Fig.~\ref{FIG:Lw} and Fig.~\ref{FIG:Delta}.

\section{Summary and Discussion}\label{sec:4}
The non-Newtonian behavior of polymeric fluids originates from the effects of polymer chain dynamics on the macroscopic rheology, which results in a non-linear relationship between shear stress and shear rate. In macroscopic continuum approaches of non-Newtonian fluids, a constitutive relation describing the non-trial dependence of shear stress on shear rate is required to close the continuum momentum equation. Rather than imposing an empirical constitutive equation, we used multiscale modeling of non-Newtonian polymeric fluids that couples macroscopic continuum equations with mesoscopic Lagrangian simulations. In particular, the equations of continuity and momentum without a closure of constitutive relation were discretized by the finite volume method~(FVM), while the dynamics of polymer chains were explicitly simulated using dissipative particle dynamics~(DPD) to provide effective constitutive closure for the continuum FVM solver.

In general, obtaining the constitutive relation covering several orders of shear rates needs a lot of individual DPD simulations at various shear rates. However, it is relatively computational expensive to perform a DPD simulation. To this end, we applied an active learning strategy to the multiscale modeling of non-Newtonian fluids to minimize the necessity of performing expensive DPD simulations. More specifically, in the multiscale simulation, the macroscopic FVM solver provides local transient flow fields to initiate DPD simulations of polymeric fluids, while the mesoscopic dynamics return the values of shear stress in terms of shear rate, wherein an active learning scheme with Gaussian process regression is used to learn the constitutive relationship from the results on a few training points. By defining proper acquisition function, the training points can be adaptively selected, so that DPD simulations are performed only when it is really necessary. Consequently, the total computational cost of the multiscale simulation is significantly reduced compared with the traditional multiscale approaches.

A non-Newtonian polymeric fluid flowing past a circular cylinder between two parallel plates at Reynolds number of 10 was used as a benchmark to confirm the effectiveness of using active learning scheme in multiscale simulations. In this case, we explicitly demonstrated how the optimal training points were adaptively selected on-the-fly informed by the FVM solver. The results showed that the multiscale simulation with active learning scheme needs only 5 DPD simulations to accurately inform a constitutive relation to close the FVM computation, where the range of shear rate was updated as he flow field evolved in time. If DPD simulations are performed at uniformly distributed shear rates in traditional multiscale approach, more than 30 individual simulations may be needed to cover the entire range of shear rates~\cite{2008Padding,2010Fedosov}. Considering that  the FVM solver is much cheaper than performing a DPD simulation, the total computational cost of the multiscale simulation is reduced by a factor of 6 by employing the proposed active learning strategy. Moreover, the active learning strategy for extrapolation always provides extra information for current needs, so that we did not need any additional DPD runs when the Reynolds number was increased from 10 to 30. When we increased the Reynolds number up to 100, only one more DPD simulation was required to get the constitutive closure for the FVM solver. This significant improvement of computational efficiency is meaningful for multiscale modeling of complex fluids whose constitutive closure needs to be computed directly from micro/mesoscale simulations.

Although we demonstrated the multiscale simulation combined with active learning scheme in polymeric fluids, the new paradigm in using machine learning tools for scale-bridging is general and can be applicable to many practical multiscale simulations of other complex fluids. We assumed that the fluid is inelastic so that a generalized Newtonian fluid model was used at the macroscale level, however,
there are more rheological properties of non-Newtonian fluids beyond the shear stress that can be obtained from DPD simulations. It would be very interesting to consider viscoelastic fluids whose constitutive closure is computed on-the-fly from mesoscopic dynamics in future work. We note that the proposed active learning strategy applied to multiscale simulation requires spatial and time separation between the continuum approach and micro/mesoscale approach, so that the message-passing multiscale modeling can be applied. Otherwise, in the cases without scale separation, the idea of applying multi-fidelity for scale-bridging~\cite{2016Babaee,2017Perdikaris} would be more useful and promising, where cross-correlations between different scales are considered. Furthermore, the computational efficiency of multiscale modeling can be greatly enhanced by applying the multi-fidelity approach as did in Ref.~\cite{2016Perdikaris}.

\section*{Acknowledgements}
This work was supported by the U.S. Army Research Laboratory and was accomplished under Cooperative Agreement No.\ W911NF-12-2-0023 and by the AFOSR Grant FA9550-17-1-0013. This work was also supported by the National Basic Research Program of China (973 Program, No.\ 2012CB025903) and NSFC~(No.\ 91434201,\ 11602133). This research was conducted using computational resources and services at the Center for Computation and Visualization, Brown University. L.\ Zhao would like to thank Dr. Yu-Hang Tang for using an in-house developed code {\em Ermine} and Dr.\ Guofei Pang for his helpful discussions on Gaussian process regression.

\end{document}